\documentclass{jetpl}
\twocolumn

\lat

\title{Anisotropy of thermal dileptons}

\rtitle{Anisotropy of thermal dileptons}

\sodtitle{Anisotropy of thermal dileptons}

\author{V.\,V.\,Goloviznin$^{a}$,
A.\,M.\,Snigirev$^{b,c}$\/\thanks{e-mail: snigirev@lav01.sinp.msu.ru},
G.\,M.\,Zinovjev$^a$}

\rauthor{V.\,V.\,Goloviznin, A.\,M.\,Snigirev, G.\,M.\,Zinovjev}

\sodauthor{Goloviznin, Snigirev, Zinovjev}

\address{$^a$ Bogolyubov Institute for Theoretical Physics, National Academy of Sciences of Ukraine, Kiev 03143, Ukraine \\
$^b$ Skobeltsyn Institute of Nuclear Physics, Lomonosov Moscow 
State University, 119991, Moscow, Russia \\
$^c$ Bogoliubov Laboratory of Theoretical Physics, JINR, 141980, Dubna, Russia }

\abstract{The meaningful specific anisotropy in the angle distribution of leptons with respect to the 
three-momentum of pair is predicted as a feasibility signature of synchrotron-like mechanism 
resulting from the quarks interacting with a collective confining color field in the heavy ion 
collisions. The lepton pair production rate and the spectrum of pair invariant mass are presented for 
this new dilepton source that is apparently not taken into consideration in the available 
phenomenological estimates.}


\begin{document}

\maketitle

\section{Introduction}
The results of experimental studies of relativistic heavy ion collisions at LHC and RHIC in recent
years have been summarized as an implication of creating hot (with high energy density) medium whose 
intrinsic degrees of freedom are the color quarks and gluons of standard quantum chromodynamics (QCD).
Alongside the compelling evidence to make reference to such an environment as a quark-gluon plasma 
(QGP) still must be found~\cite{jacmuel, schukr} and well elaborated. The electromagnetic probes
(photons and dileptons), long been proposed with this goal~\cite{feinberg,shuryak}, still remain one 
of the most informative, even when their momenta are not very large, because their weak electromagnetic 
interactions provide them with a mean free path inside medium essentially larger than the medium size.
Photons and leptons are freely emitted from this excited region, practically without interaction with 
the color quarks and gluons, and their undistored spectra are carrying a direct information on the 
state of hot excited medium~\cite{Shen:2016odt,Paquet:2016pnt}.

In this context the recent measurements by the PHENIX Collaboration which show the azimuthal 
anisotropy of produced direct photons very close to the hadron one~\cite{phenix2011} are rather 
exciting. This result appears to be in a serious contradiction with expected dominance of photon 
production from QGP at an early stage of ion collision at the top RHIC and now available LHC energies.
The observed temperature of "anomalous" photon radiation (about $T_{\rm ave}\simeq220$ Mev) is in 
accordance with the PHENIX Collaboration measurements~\cite{phenix2010} at the energy 
$\sqrt{s}=200$ GeV of heavy ion collisions. This temperature magnitude being considered as a result of
averaging over the entire evolution of the matter created in nuclear collisions is noticeably higher 
than the expected critical temperature~\cite{campbell} and obviously supports the scenario of photon 
radiation from QGP. Clearly, such a situation is nontrivial for the phenomenological studies because the 
production rate of real and virtual (low mass dileptons) generated by a hot QGP is considered 
increasing like $T^4$ and, hence, being very sensitive to the temperature of medium. Fortunately, the 
situation with improving a description of global photon data overall is gradually becoming more 
controlled~\cite{vujanovic} (at least with one disadvantage which is a number of photon sources getting 
larger).

Recent interesting suggestions for photon and dilepton sources~\cite{Chatterjee:2005de} (thermal photons may have elliptic flow), \cite{pisarski} (thermal radiation from 
semi-QGP), and~\cite{chiu} (forming a gluon condensate that radiates the photons at the early stage of 
collisions) which are idealogically close to the scinario we develop in this letter declare pretty small
photon azimuthal anisotropy~\cite{chatterjee} and insufficient to explain the experimental data 
mentioned. There are many other phenomenological models under discussion (see, for example, ~\cite{hees,
kharzeev,bzdak,liu,linnyk,Zakharov:2016mmc,Zakharov:2017cul}) which are in different extent fairly 
successful at treating the experimental data quantitatively but sometimes with noticeable uncertainties. 
In our previous work~\cite{Goloviznin:2012dy} we have suggested significantly alternative mechanism 
that contributes to the observed anisotropy of direct photons. The reference is to a "magnetic 
bremsstrahlung-like radiation" (or synchrotron radiation in present terminology) of quarks in the 
collective color field ensuring confinement. We have found that this boundary bremsstrahlung is
intensive enough~\cite{gol1,gol2,gol3}, develops the azimuthal anisotropy~\cite{Goloviznin:2012dy} 
and is capable of resolving the "direct photons puzzle"~\cite{phenix2011} still without appealing 
to the non-equilibrium dynamics of heavy ion collision process.

The main goal of our present letter is to show that the mechanism above predicts also the anisotropy of
dileptons that now is considered~\cite{Baym:2017qxy,Baym1} as a good probe of QGP and effective instrument
for resolving the discussed discrepancy between the experimentally observed dilepton spectra and the 
theoretical expectations~\cite{Shen:2016odt,Paquet:2016pnt,vujanovic,pisarski}. In Sec. II we give the 
basic equations and estimates of the total number of lepton pairs and their spectra. The peculiarity 
in the angular dilepton distribution is discussed in Sec. III. In the conclusion we summarize the main 
results. 
 
\section{Dilepton spectra}

An existence of the boundary bremsstrahlung is based on three quite realistic assumptions: 1) the 
presence of relativistic light quarks ($u$ and $d$ quarks) in the hot medium; 2) the semiclassical 
nature of their motion; 3) confinement. Then as a result, each quark (antiquark) at the boundary of the system volume 
moves along a curve trajectory and (as any classical charge undergoes an acceleration) emits photons. 
Estimating the magnitude of this effect we have utilized~\cite{gol1,gol2,gol3} the chromoelectric flux 
tube model ~\cite{tube1,tube2,tube3} in which the interaction between the volume of quark-gluon system 
and color object crossing over its boundary develops the constant force $\sigma$ bringing a color object 
back. Apparently, this force is acting along the normal to the plasma surface. Quantitatively, an effect 
is rooted in the large magnitude of quark confining force $\sigma \simeq 0.2$~Gev$^2$. It is easy to 
recognize that this mechanism could be an alternative one for generating the lepton pairs, too, as it 
has been argued in our old paper~\cite{gol4}, some results from that we reproduce below to be clear.

A large value of $\sigma$ results in the large magnitude of characteristic parameter 
$\chi = ((3/2) \sigma E/m^3)^{1/3}$ (where $E$ and $m$ are the energy and mass of the emitting 
particle, respectively) for $u$ and $d$ quarks (the strong-field case). In this regime the 
probability of emitting a "massive" photon is independent of the mass of the emitting particle and 
in the first order in inverse powers of the parameter $\chi$ can be written as~\cite{zhulego}
\begin{equation}
\label{d1}
dW_{\gamma}(M^2) /dt = 1.56 e_q^2 \alpha (\sigma \sin \varphi)^{2/3} E^{-1/3}, 
\end{equation}
where $\alpha=1/137$ is the fine structure constant, $e_q$ is the quark charge in units of electron 
charge and $\varphi$ is the angle between the quark velocity and the direction of quark confining 
force (the normal to the QGP surface in our case). Using the well-known relation between the cross 
sections for virtual-photon and lepton-pair production, from Eq.~(\ref{d1}) we easily find the 
lepton-pair distribution in the invariant mass: 
\begin{eqnarray}
\label{d2}
& &\frac{dN}{dtdM^2}  =   
\frac{\alpha}{3\pi} f(M)\frac{dW_{\gamma}(M^2)}{dt},\\
& & f(M) = \frac{1}{M^2}\Bigg(1+\frac{2\mu^2}{M^2}\Bigg)
\Bigg(1-\frac{4\mu^2}{M^2}\Bigg)^{1/2}, \nonumber \\
& & 2\mu \leq M \leq E. \nonumber 
\end{eqnarray}
In this equation $\mu$  and $M$ stand for the lepton mass and the invariant mass of the pair, 
respectively. Both these equations are invalid only for the invariant masses $M$ close to $E$. 

Further, in order to obtain the number of lepton pairs radiated per unit surface area of QGP per unit 
time in invariant mass interval $M^2$, $M^2 + dM^2$, it is necessary to average Eq.~(\ref{d2}) over the 
quark paths and to convolute it with the flux of quarks reaching the boundary of the QGP volume from 
within. This procedure does not differ from the analogous one performed in detail in 
Refs.~\cite{gol2,gol3} for photons spectra, so we present only the final result here:
\begin{eqnarray}
\label{d3}
\frac{dN}{dSdtdM^2} & = &A \alpha^2 \sigma^{-1/3} f(M)M^{11/3} \\
& & \times \int\limits_1^{\infty} d\xi (\xi^{8/3} -1) \exp\Bigg(-\frac{M\xi}{T}\Bigg),\nonumber
\end{eqnarray}
where
$$A = \frac{1.56}{2(2\pi)^3}\frac{\Gamma(4/3)\Gamma(1/2)}{\Gamma(11/6)}g(e_u^2
+e_d^2),$$ 
$\Gamma$ is the gamma-function, $e_u$ and $ e_d$ are the $u$- and $ d$-quark charges, 
$g =$ spin $\times$ color = 6 is the number of quark degrees of freedom, $T$ is the plasma temperature. 
Integrating Eq.~(\ref{d3}) over $dM^2$ one obtains the total number of lepton pairs emitted per unit 
time from unit surface area of QGP as
\begin{eqnarray}
\label{d4}
\frac{dN}{dSdt} &= &\frac{16}{3} A\alpha^2
\sigma^{-1/3}T^{11/3}\int\limits_{\beta}^{\infty} dye^{-y}y^{5/3} \\
& & \times\Bigg\{ \ln
\Bigg[\frac{y}{\beta} \Bigg(1+\Bigg(1-\frac{\beta^2}{y^2}\Bigg)^{1/2}\Bigg)\Bigg] -
\nonumber \\
& & -\frac{1}{6}\Bigg(1-\frac{\beta^2}{y^2}\Bigg)^{1/2}\Bigg(5+\frac{\beta^2}{y^2}
\Bigg)\Bigg\},\nonumber
\end{eqnarray}
where $\beta = 2\mu/T$. At $\beta \ll 1$ Eq.~(\ref{d4}) is essentially simplified
$$\frac{dN}{dSdt} \simeq  2A \Gamma \Bigg(\frac{11}{3}\Bigg) \alpha^2
\sigma^{-1/3} T^{11/3} [\ln(T/2\mu )+a +O(\beta^2)],$$
\begin{equation}
\label{d5}
~ 
\end{equation}
$$a = \ln2-5/6 + \Gamma{'}(8/3) /\Gamma(8/3).$$
This is a reasonable estimate of the total number of electron-positron pairs, since $\mu_e \simeq $ 0.5
MeV is considerably less than the minimal plasma temperature $T \simeq $ 200 MeV.

In the simplest case, if the plasma occupies a spherical volume of radius $R$ and does exist during the 
time $\tau$, then the total number of electron-positron pairs is easy estimated as
\begin{equation}
\label{d6}
N=4\pi R^2 \tau dN/dSdt.
\end{equation}
Of course, it is interesting to compare this result with the total number of electron-positron pairs 
produced by "standard" quark-antiquark annihilation processes in the QGP volume
~\cite{feinberg, shuryak, kajantie} 
\begin{equation}
\label{d7}
N_{ann} = \frac{4}{3} \pi R^3 \tau B \alpha^2 T^4,~~~~~B=10/9\pi^3,
\end{equation}
which takes into account only the  $u$- and $d$-quark contributions to the rate to make 
comparison adequately here. There are other QGP volume (and not QGP) contributions except the Born term (see, e.g.,~\cite{Altherr:1992ti,Roy:1996gt}) which should be taken into consideration at detail analysis, especially for lower invariant masses.  Then the relevant quantity is the ratio
\begin{equation}
\label{d8}
\frac{N}{N_{ann}} = \frac{C}{RT^{1/3}\sigma^{1/3}}
\Bigg(\ln\frac{T}{2\mu}+a\Bigg),
\end{equation}
where $C=6\Gamma(11/3)A/B \approx 11.8$. Numerically $N/N_{ann}\simeq $40 on setting $R=1$ fm , and 
$N/N_{ann}\simeq $4 on setting $R=10$ fm at $T\simeq200$ MeV. Therefore, for QGP systems of the 
expected size (5-10) fm, the mechanism outlined above contributes dominantly to the total number of 
electron-positron pairs produced by the plasma. 

This result is still valid when the space-time plasma evolution has been included 
following~\cite{bjorken}. Indeed, the corresponding integration over $dSdt$ and $d^4x$ can be performed
as in Ref.~\cite{gol3} if one neglects additional logarithmic dependence on the temperature in 
Eq.~(\ref{d5}), taking $\ln(T_c/2\mu)$ ($T_c$ is the phase transition temperature) instead of 
$\ln(T/2\mu)$. Then, as for the photons~\cite{gol3}, the functional distinction between the proposed 
mechanism and the "standard" volumetric one is mainly determined by the parameter that is just the 
dimensionless combination as
 $$ (RT^{1/3}_c\sigma^{1/3})^{-1}$$
(note that it is $\simeq 1$ on setting $R\simeq 0.6$ fm) with a constant which is slightly different 
from $C [\ln(T_c/2\mu )+a]$) incoming in Eq.~(\ref{d8}).

For muon pairs, the ratio $2\mu/T\simeq 1$ and the integration over invariant mass cannot be performed 
analytically. However, on extracting the basis temperature dependence $T^{11/3}$, one can estimate 
numerically the remaining integral $I$ in Eq.~(\ref{d4}) since it is a slowly varying function over the 
small temperature interval (in the models based on the scaling solution of hydrodynamical equations 
$T=200-500$ MeV).Thus the ratio, analogous to Eq.~(\ref{d8}) for the muon-pair production from the 
plasma of a spherical volume with no space-time evolution, is
\begin{equation}
\label{d9}
(N/N_{an})_{muon}=C_{muon}/RT^{1/3}\sigma^{1/3},
\end{equation}
where $C_{muon}\simeq 7.1$, demonstrating the significant contribution of the bremsstrahlung mechanism. 
The ratio of the invariant mass spectra is also analytically estimated in the regime $T \ll M$~
\cite{gol4} with the similar conclusion.

\section{Angular distribution}

One of the most distinctive features of the proposed mechanism is a large degree of photon 
polarization~\cite{gol2, gol3}. For a plasma of cylindrical symmetric volume with its axis along the 
collision axis, the bremsstrahlung photons are dominantly polarized along the normal to the plane 
spanned by the cylinder axis and the momentum of registered photons. The appearance of such a 
polarization is closely connected with the choice of direction of the collective field where quarks are 
moving and its value is virtually insensitive to the parameter regulating an intensity of bremsstahlung.
Consequently, these photons can also be polarized for other "nonideal" shapes of plasma surface 
possessing this decisive property. The corresponding calculations of the polarization degree in a 
lucid form can be very complicated.

We realize that the difficulties in registering photon polarization entail many problems for 
experimental search for this effect. But observing lepton-pair spectra resulting from the polarization 
of intermediate photon could be a potentially efficient probe of QGP~\cite{Baym:2017qxy,Baym1} 
if formed in collisions of ultrarelativistic ions.

Considering the decay of massive photons with the four-momentum k into a lepton pair, the following 
expression gives the squared matrix element of this process:
\begin{eqnarray}
\label{f1}
|M|^2 = 4\pi \alpha Sp[(\hat p_1 + \mu) \gamma_{\mu} (\hat p_2 -
\mu)\gamma_{\nu}]e_{\mu}e_{\nu}^* = \nonumber \\
= 16\pi \alpha [k^2/2 + (p_1e)(p_2e^*) + (p_1e^*)(p_2e)],
\end{eqnarray}
where $e$ is the polarization four-vector of the photon and  $(ee^*) = -1$;  $p_1$ and  $p_2$  are the 
four-momenta of the lepton and antilepton, respectively.


Drawing the relevant phase space of the pair and taking into account the transversality condition 
$(ek)=0$, the lepton distribution per unit time in the radiation angle reads as
\begin{eqnarray}
\label{f3}
\frac{dW}{dtd\Omega_1} & = &\frac{\alpha}{2\pi k^0}\int \frac{p^2dp}{p_1^0(k^0-p_1^0)}\delta[f(p)] \nonumber\\
& & \times [k^2/2 - 2(p_1e)(p_1e^*)],
\end{eqnarray}
where
$$f(p) = k^0 - p_1^0 - (\mu^2 + {\bf k}^2 + p^2 -2|{\bf k}|p \cos \theta_1)
^{1/2}.$$ 
Here $\theta_1$ is the angle between the three-vectors ${\bf k}$ and ${\bf p_1}$ and the condition 
$f(p)=0$ determines the length of the three-vector ${\bf p_1}$ as a function of $\cos \theta_1$.

If the initial photons are unpolarized, Eq.(\ref{f3}) has to be averaged over polarization and then 
it results to the lepton distribution independent of the radiation azimuthal angle $\phi_1$. This 
dependence exists at decay of the polarized photons. Defining $ n(1+\delta)/3$ as the photon number 
of the states with polarization vector $e_1$, $ n(1-\delta)/3$ as the photon number of the states with 
polarization vector $e_2$ and $ n/3$ as the same with polarization vector $e_3$, and choosing the 
reference frame with the $z$ axis directed along the three-vector ${\bf k}$ and the $x$ and $y$ axes 
tallying with the directions of ${\bf e_1}$ and ${\bf e_2}$, we have then 
$$e_1=\{0,~1,~0,~0\},~e_2=\{0,~0,~1,~0\},$$
$$e_3=\{|{\bf k}|/\sqrt{k^2},~0,~0,~k^0/\sqrt{k^2}\},
~k=\{k^0,~0,~0,~|{\bf k}|\},$$ 
$$p_1=\{\sqrt{p^2+\mu^2},~ p\sin\theta_1 \cos
\phi_1,~p\sin\theta_1  \sin\phi_1,~p\cos\theta_1\}.$$
Finally, the lepton distribution in the radiation angle takes the form
\begin{eqnarray}
\label{f4}
\frac{dN}{dtd\Omega_1} & = &\frac{\alpha n}{2\pi k^0} \int \frac{p^2dp}{p_1^0 (k^0 - p_1^0)} \delta[f(p)]\\
& & \times  \Bigg[\frac{k^2+2\mu^2}{3}
- \frac{2}{3} \delta p^2 \sin^2 \theta_1
\cos 2\phi_1
\Bigg].\nonumber
\end{eqnarray}
It follows from Eq.~(\ref{f4}) that the angular lepton distribution has a characteristic dependence on 
the azimuthal angle $\phi_1$ if a massive photon has transverse (in the three-dimensional space) 
polarization ($\delta$ is not zero). When the photons are unpolarized or have longitudinal polarization
(along the vector $e_3$) this dependence disappears. Thus the azimuthal anisotropy of the radiation 
angle distribution of leptons certainly indicates that the intermediate ("massive") photon has a 
transverse polarization. We may argue this effect as a highly feasible to be experimentally measured 
and considered as a probe of QGP events just in accordance with the recent consideration~\cite{Baym:2017qxy, Baym1}.

Indeed, it was shown \cite{gol3} that for a plasma with a cylindrically symmetric volume the value of 
$\delta$ can be calculated as about 20 $\%$. In our case, the intermediate photons (as we have a strong
field regime) could be considered up to the masses $\sqrt{k^2} \simeq \sqrt{\sigma}=0.45$ GeV as having
a small virtuality and their properties are quite close to real photons~\cite{zhulego}. It means these 
photons are transversely polarized with practically the same degree of polarization $\delta$. Since as 
the proposed mechanism contributes noticeably to the total yield of lepton pairs from QGP, the 
"bremsstrahlung" leptons could be identified by measuring their angle anisotropy that is absent in the 
Drell-Yan mechanism and the "standard" volumetric mechanism.

\section{Conclusions}

Our analysis shows that the interaction of quarks with the collective color field confining them results 
in an intensive radiation of the magnetic bremsstrahlung type (synchrotron radiation). The intensity of
such a radiation for the hot medium of size 1-10 fm that is expected in ultrarelativistic collisions 
of heavy ions is comparable with that of the volume mechanism of photon and dilepton production in
the temperature range of $T= 200-500$ MeV. Quantitavelly a relative effect is regulated by the three basic 
parameters: the characteristic medium (QGP) size $R$, the QGP temperature $T$, and the confining force 
$\sigma$, which are firmly fixed. Possible uncertainties come mainly from the simple modeling of 
confinement and simplification of the QGP geometry what allow us to obtain estimates in transparent
analytical form.

The most striking feature of magnetic bremsstrahlung is the high degree ($\sim 20 \%$) of polarization 
of both real and "massive" (virtual) photons that is mainly determined by the medium (QGP) geometry. The 
virtual photons develop the noticeable specific anisotropy in the angle distribution of leptons with 
respect to the three-momentum of pair. The origin of this anisotropy is rooted in the existence of a 
characteristic direction in the field where the quarks are moving. Besides the synchrotron radiation  
will be nonisotropic~\cite{Goloviznin:2012dy} for the noncentral collisions because the photons are 
dominantly emitted around the direction fixed by a surface normal. As result the coefficient of 
elliptic anisotropy for dilepton pairs (the study of which was suggested in Ref.~\cite{chatterjee}) will be also proportional to the eccentricity of QGP system as it takes place for the bremsstrahlung real photons and can be experimentally measured.

Indeed, in order to draw a more definite conclusion, further investigations are necessary including, in particularly, a proper comparison with other sources of photons and dileptons.


The papers by G. Baym in this field were quite instructive and 
imaginative for us since the first discussion of one of us (GMZ) with him 
at the 6th Quark Matter in Nordkirchen (1987). Critical remarks by B. Zakharov were quite instructive. Several comprehensive discussions of these results with V. Toneev are gratefully acknowledged. The paper was partially supported by the Goal-Oriented Program of Cooperation between CERN, JINR and National Academy of Science of Ukraine "Nuclear Matter under Extreme Conditions".



%

\end{document}